\documentstyle[11pt,aaspp4]{article}
\def\vi{\hbox{$V\!-\!I$}}
\def\bi{\hbox{$B\!-\!I$}}

\begin{document}

\title{Gas-Rich Dwarf Galaxies from the PSS-II --- II: Optical Properties}

\author{Rachel A. Pildis\altaffilmark{1}}
\affil{Department of Astronomy, University of Michigan; and
\\ Harvard-Smithsonian Center for Astrophysics, 60 Garden
St., MS 83,\\Cambridge, Massachusetts  02138;\\ r-pildis@nwu.edu}
\altaffiltext{1}{Present address: Dearborn Observatory, Northwestern
University, 2131 Sheridan Road, Evanston, Illinois  60208-2900}

\author{James M. Schombert}
\affil{Department of Physics, University of Oregon, 120 Willamette Hall,
Eugene, Oregon  97403;\\ js@abyss.uoregon.edu}

\and

\author{Jo Ann Eder}
\affil{Arecibo Observatory\altaffilmark{2}, P.O. Box 995,
Arecibo, Puerto Rico  00613;\\ eder@naic.edu}
\altaffiltext{2}{Arecibo Observatory is part of the National Astronomy and
Ionosphere Center, which is operated by Cornell University under contract
with the National Science Foundation.}

\slugcomment{To appear in The Astrophysical Journal, 20 May 1996}

\begin{abstract}

We describe the optical properties of a sample of 101 gas-rich field dwarf
galaxies found on PSS-II (Second Palomar Sky Survey) plates, most newly
discovered as part of a survey to investigate the clustering properties of
dwarf galaxies relative to giants.  These galaxies have low surface
brightnesses and are relatively distant, with recession velocities ranging
up to $10^4$ km s$^{-1}$.  They have bluer \vi\ colors (median value of
0.75) than either actively star-forming giant galaxies or low metallicity
globular clusters, implying that these dwarfs have both low metallicities
and little past star formation.  These galaxies are also extremely gas rich,
with a median HI mass to $V$ luminosity ratio of approximately 2 in solar
units.  We divide the sample into two groups:  true dwarfs with diameters
(at 25 $I$ mag arcsec$^{-2}$) less than 7.5 kpc and Magellanic dwarfs with
diameters greater than that value.  The true dwarfs have greater HI mass to
$V$ luminosity ratios and slightly bluer \vi\ colors than the Magellanic
dwarfs.  Overall, the optical properties of our sample of dwarf galaxies
point towards their being quiescent objects that have undergone little star
formation over the age of the universe.  They are not faded objects, but
instead may be going through one of their first periods of weak star
formation.

\end{abstract}


\pagebreak

\section{Introduction}

	Dwarf galaxies, while they are considerably smaller and less
luminous than ``normal'' galaxies such as the Milky Way, provide important
clues to our understanding of the universe.  On the largest scales, dwarfs
have been invoked as a test of possible biasing between the distribution of
dark matter and that of bright galaxies (e.g., \cite{dek86}) and as possible
sites of extremely high mass-to-light ratios (e.g., \cite{lak90}).  On much
smaller scales, dwarfs and other low-surface-brightness (LSB) galaxies
often are examples of star formation in nearly pristine low-metallicity
environments and thus aid in the understanding of the variation of stellar
initial mass functions and the evolution of metal enrichment in metal-poor
conditions (e.g., \cite{deh94,mcg94a,mar95}).  Certainly, an understanding
of galaxies as a whole must include an explanation for the formation and
evolution of dwarfs as well as giants.

	We have undertaken a survey of dwarf galaxies discovered in a visual
search of 1400 deg$^2$ of plate material from the Second Palomar Sky
Survey (PSS-II).  Our catalog contains 145 objects detected in neutral hydrogen
emission with the Arecibo 305 m telescope, from an initial list of 350
candidates (\cite{sch97}, Paper I in this series).  These objects tend to have
low surface brightnesses even though this characteristic was not one of
the selection criteria (\cite{sch97}).  HI detections are
required in order to determine a redshift distance for these small and faint
objects.  This survey is an outgrowth of an earlier investigation of dwarf
gas-rich galaxies (\cite{ede89}).  The main motivation for that survey and
this current one is to test the prediction of biased galaxy formation that
dwarfs trace the dark matter more closely than do bright galaxies, and thus
that dwarfs cluster less tightly. This current survey covers far more of the
sky than that of Eder et al.\ (1989), and this paper focuses on an aspect of
dwarf galaxies unexplored by that earlier work:  their optical properties.

	Recent investigations of giant LSB galaxies have shown that they
have average optical colors considerably bluer than high-surface-brightness
galaxies of the same size.  McGaugh \& Bothun (1994) found that giant LSB
galaxies are even bluer than actively star-forming Sc galaxies, even though
little active star formation is seen in LSBs.  Comparisons with
low-metallicity globular clusters show that extremely low metallicities
alone cannot explain these blue colors, implying that some combination of
low metallicity and recent star formation is needed (\cite{mcg94b,deb95}).
A study of extremely large LSB galaxies concluded that LSB galaxies with
larger disk scale lengths are redder than those with smaller scale lengths,
a trend which is similar to what is seen in high-surface-brightness galaxies
(\cite{spr95}).

	This paper will discuss the optical observations and data reduction
for our sample of galaxies (\S 2) and the initial analysis done to obtain
the optical properties of this sample (\S 3).  Section 4 lists the criteria
used to determine our final sample of dwarf galaxies and discusses how these
properties correlate with one another and with the HI properties detailed in
an upcoming paper (\cite{ede97}, Paper III in this series).  The final
section (\S 5) summarizes our results and discusses their implications.

\section{Observations and Reduction}

	Optical images were obtained under photometric conditions during
1992, 1993, and 1994, using the 2.4m Hiltner telescope at the
Michigan-Dartmouth-M.I.T. (MDM) Observatory on Kitt Peak, Arizona.  Details
of the individual runs are given in Table 1.  We increased the pixel size
over the course of the observations in order to minimize both noise and
readout time; since this study concentrates on the overall structure and
color of these dwarf galaxies and not fine detail within them, the small
loss of resolution is unimportant.

	Each galaxy was imaged in $I$ for a minimum of 15 minutes, and in
$V$ for a minimum of 25 minutes, with individual exposures lasting 300
seconds each.  Photometric calibrations were made using observations of
standard stars from Landolt (1992) and the standard Kitt Peak airmass
coefficients.  Raw color measurements are accurate to $\sim$0.05 magnitudes
and the error in the zero point is $\sim$0.03 in both filters.  Corrections
for galactic absorption for the total $V$ magnitudes and average \vi\ colors
were made using the formula of Sandage (1973).  Most of the galaxies have
negligible galactic absorption: only four (D702-1, D774-1, D774-2, and
D774-3) have $V$ magnitude corrections greater than 0.2 (ranging from 0.26
to 0.36) and \vi\ corrections greater than 0.1 magnitudes (0.13--0.18).

	The raw data were turned into coadded images of each galaxy in the
$V$ and $I$ bands using utilities in IRAF.  After the average bias level and
the average bias structure were subtracted from all frames, flat fields were
created from twilight exposures (for Thomson CCD images) or object frames
(for Loral CCD images).  For the latter, the location of targeted galaxies
on the CCD was changed for each pointing in order to make more accurate
flats.  The resulting flat fields were applied to each object frame before
the frames were coadded to make a summed image of each galaxy in each of
the two filters used.

\section{Initial Analysis}

	The photometry for this project was performed using elliptical
apertures based on the surface photometry of the individual galaxies.  The
surface photometry was determined using ARCHANGEL (\cite{sch89}), a
derivative of the GASP package of galaxy surface photometry routines
supplemented by various other programs written by the authors.  These
algorithms were used to transform the 2D instrumental CCD frames to 1D
calibrated surface photometry.

	The first step in this process is the determination of the sky value
and its error. This was done by displaying the image and interactively
selecting fifteen to twenty regions located in a symmetrical pattern around
the object of interest while avoiding low surface brightness features,
halos of bright stars, or any other objects that might degrade the sky
estimate.  The size of the region can be varied, but was usually a 15 by 15
box. The pixels inside the box are averaged using a 3$\sigma$ clipping to
remove cosmic rays.  The means of the boxes are then averaged for the final
sky value.

	The second step is to remove objects from the frame unrelated to the
dwarf galaxies that may confuse or distort the routines that fit ellipses of
constant surface brightness.  This was also done interactively to produce a
mask of stars, background galaxies, and bad columns.  The masked pixels are
set to a value that is ignored by the software (rather than attempting to
interpolate over the masked regions) and the mask is applied to the image in
each filter in a uniform fashion.  This has the consequence of removing
luminosity from the galaxy data, but the number of masked pixels in the
galaxy is usually less than 2\% and primarily located in the outer faint
envelopes.  The impact on the total luminosity is negligible.  Since the
colors are determined from identically masked frames, there is no impact on
aperture or isophotal colors.

	The ellipse fitting to determine the surface photometry was done
using the PROF routine in the GASP package. A full description appears in
Davis et al.~(1985), but in brief, PROF is an iterative least-squares
procedure adjusting the coefficients of a Fourier series that describe the
variation of intensity values in a sampling ellipse.  After a threshold is
reached, or a maximum index is exceeded, the sampling region is moved
outward and another fit is started.  All ellipse parameters are fitted (not
fixed), and the increase in ellipse radius for each successive fit is set at
10\%.  The output from the program is a file containing the ellipse
parameters (center, major axis, eccentricity, position angle),
goodness-of-fit criteria, and the mean intensity (in counts per pixel) of
the ellipse.  Final profiles are stored as major axis in arcsecs versus
surface brightness.  Scale lengths are then derived from least-squares
fitting of an exponential function to the data in the surface
brightness--radius plane.

	The resulting ellipses from the surface photometry are then used to
perform elliptical aperture photometry to determine total magnitudes and
colors.  Total magnitudes are determined by a curve-of-growth method where
the outermost fitted ellipse is used to follow the galaxy's light to the
edge of the CCD frame. We used the $I$ band data to produce the ellipses and
resulting surface photometry since the shapes of the galaxies are smoother
in $I$ than $V$.  The $I$ band ellipses were applied to the $V$ data to
obtain aperture magnitudes.

	The total colors and color profiles were obtained from the
elliptical aperture photometry.  Since all the data were registered and
masked using the same templates, the color measurements thus are ensured
to use the same pixels from frame to frame.  Aperture colors are simply the
difference of the aperture magnitudes of the bandpasses of interest.  In
addition, differential colors were produced by comparing the luminosities in
annular rings based on the surface photometry ellipses.  Selected integrated
color profiles and differential color profiles are shown in Fig.\ 1, and
profiles for the entire sample will be available on the World Wide
Web\footnotemark.
\footnotetext{http://zebu.uoregon.edu/$\sim$js/dwarf.html}

	A summary of the basic properties of all the galaxies with both
photometric optical data and unconfused HI detections is given in Table 2.
The HI data are taken from Eder et al.\ (1997), or from the NASA/IPAC
Extragalactic Database if a galaxy was known to have a previous HI
detection.  Included in this table are each galaxy's name (column 1),
heliocentric velocity measured in HI (column 2), absolute $V$ magnitude
(column 3), log of its neutral hydrogen mass in solar masses (column 4),
radius at the 25 $I$ magnitude arcsec$^{-2}$ isophote (column 5---since LSB
galaxies are observed to have \bi=1.2--1.5, this is approximately equivalent
to a Holmberg radius), average \vi\ color (column 6), scale length (column
7), central surface brightness (column 8), and other names of the galaxy
(column 9, which also marks those galaxies determined to be dwarf spirals in
\cite{sch95}).  For all distance-related values in this paper, we assume a
Hubble constant H$_0$=85 km s$^{-1}$ Mpc$^{-1}$, $\Omega_0$=0.2, and a
Virgocentric infall model with $v_{infall}$=300 km s$^{-1}$ and
D$_{Virgo}$=15 Mpc.  The mean colors quoted in Table 2 are based on the
averaged colors of the differential profiles weighted by the surface
brightness of the annulus.  As discussed below, a few of the galaxies listed
in Table 2 are too luminous to be classified as dwarfs or have highly
uncertain distances, and thus will be excluded from further analysis.

\section{Results}

	In this paper we define our final sample as those objects that have
both an absolute $V$ magnitude fainter than $-19$ and a heliocentric
velocity of greater than 500 km s$^{-1}$.  The second restriction is needed
due to the large uncertainty in calculating distances for low velocity
galaxies.  These restrictions reduce the number of galaxies in our sample
from 109 to 101.

	We further subdivide the final sample into two categories based on
the diameter D$_{25}$=$2 \times$R$_{25}$.  The larger of the two groups (55
galaxies out of 101) have D$_{25}$ $<$ 7.5 kpc (R$_{25}$ $<$ 3.75 kpc) and
will be designated as ``true dwarfs''.  The galaxies with D$_{25}$ $>$ 7.5
kpc will be called ``Magellanic dwarfs'', since they are roughly the size
and luminosity of the Large Magellanic Cloud.  This division was chosen
because it was the best size-based discriminator to separate out those
galaxies that are the brightest (-17 $>$ M$_V$ $>$ -19) and contain the most
mass in HI (M$_{HI}$ $>$ $10^9$ M$_{\sun}$) from their smaller cousins.
This can be seen most clearly in Fig.\ 2.  Figure 3 shows that this division
also separates the galaxies with scale lengths of roughly 1 kpc or less from
more extended objects.

	Figure 2 also shows that these galaxies, as expected, contain a
great deal of neutral gas, with most having an HI mass to $V$ luminosity
ratio (where both quantities are in solar units) of unity or greater.  The
median value of this ratio is 2.4 for the true dwarfs and 1.9 for the
Magellanic dwarfs, and 2.0 for the sample as a whole.  A more common ratio
used in the literature is HI mass to $B$ luminosity:  for \bv=0.5, a typical
color for LSB galaxies (\cite{deb95}), the median values of that quantity for
the two subsamples are 2.1 and 1.7, respectively.  Median HI mass to $B$
luminosity ratios for UGC spiral galaxies range from 0.1 to 0.5, with
irregulars having slightly higher values (\cite{rob94}).  van Zee, Haynes,
\& Giovanelli (1995) found that the median HI mass to $B$ luminosity ratio
for a large sample of dwarfs was 0.9, with fewer than 15\% of the galaxies
having a ratio above 5.  Clearly, our sample of dwarf galaxies is extremely
gas-rich or, alternatively, quite underluminous for their gas masses.
This should not be construed to mean that all dwarf galaxies are gas-rich;
our requirement of an HI detection to determine a redshift for each object
eliminates gas-poor objects from our sample.

	The sample galaxies have very blue average \vi\ colors, with a
median color of \vi=0.75 (Fig.\ 4).  This is bluer than the \vi\ values of
0.85--0.90 seen in both high-surface-brightness extreme late-type galaxies
(\cite{deb95}) and the most metal-poor globular clusters
(\cite{hes85,ree88}), and comparable to the values found for for large LSB
galaxies (\vi=0.70--0.95; \cite{mcg94b,deb95}).  Furthermore, standard stellar
synthesis models only produce \vi\ values of under 0.9 with when little past
star formation is assumed (\cite{gui87,maz92}).

	The \vi\ color of a galaxy is sensitive to both the position and
degree of development of the giant branch, with blue colors indicating that
the giant branch is low in metallicity or underpopulated or both
(\cite{mcg94b}).  The very blue \vi\ colors of the galaxies in our sample,
along with their large HI mass to optical luminosity ratios, argue that
these are metal-poor galaxies with only sporadic star formation in the
past.  The median color for true dwarfs is somewhat (0.11 mag) bluer than
that of the Magellanic dwarfs, perhaps indicating that the true dwarfs have
had more recent star formation or that they are more metal-poor, or some
combination of the two.  These colors follow the trends seen in other
studies of LSB galaxies:  LSB galaxies are bluer than
high-surface-brightness galaxies of the same type, but the larger LSB
galaxies are somewhat redder than ones of smaller size
(\cite{mcg94b,deb95,spr95}).

	Figure 5 is a color-magnitude diagram for our dwarf sample.  Only a
very small trend for the more luminous galaxies to be redder than the dimmer
ones can be seen.  Even at the highest luminosities, however, the galaxies
in our sample tend to be bluer than the average \vi\ color of 0.9 seen in
late-type galaxies with high surface brightnesses (\cite{deb95}).  Again,
this points toward LSB galaxies having less past star formation than their
high-surface-brightness cousins, leading to lower metallicities and less gas
depletion.  Low surface densities of neutral hydrogen may be a cause of this
relative lack of past star formation, as seems to be the case in large LSB
galaxies (e.g., \cite{van93}).  Mapping of the HI surface density and
velocity structure in a subset of our sample is currently underway.

	As is seen in LSB galaxies of all sizes (\cite{mcg94b,spr95}), there
is no clear correlation between the disk parameters $\mu_0$, the central
surface brightness of the disk (here in $I$ magnitudes arcsec$^{-2}$) and
$\alpha$, the disk scale length (Fig.\ 6).  The empty area in the lower left
hand corner of the plot is simply our selection bias against very small,
very faint objects (i.e., such objects are surpassingly difficult to find on
sky survey plates).  Most of the galaxies in our sample have central surface
brightnesses of 21--23 $I$ mag arcsec$^{-2}$, which for their median
\vi\ color of 0.75 and average LSB \bv\ color of 0.5 (\cite{deb95}) are
equivalent to $B$ band surface brightnesses of 22.3--24.3.  This is
similar to the values of $\mu_{0,B}$ seen in
surveys of large to giant LSB galaxies (\cite{mcg94b,spr95,deb95}).

	A comparison of the disk parameters $\alpha$ and $\mu_0$ with the
average \vi\ color of the galaxies reveals only slightly more of a trend
than is seen in comparing the parameters to each other (Fig.\ 7).  Since the
smaller galaxies in this sample (i.e., the true dwarfs) are somewhat bluer
than the larger ones and our diameter criterion correlates well with the
disk scale length, it is not surprising to see in Fig.\ 7a that $<$\vi$>$
becomes smaller with decreasing $\alpha$.  Note, however, that this is true
only on average; the bluest galaxies are not the smallest in scale length,
and vice versa.  The trend, however, is stronger than that seen in Fig.\ 5
for average \vi\ color versus absolute $V$ magnitude, which implies that
scale length may be a better predictor of galaxy properties than luminosity
is.

	Figure 7b reveals little correlation between the disk central
surface brightness and the overall color of a galaxy, with a small
exception.  The three true dwarfs with the lowest surface brightnesses also
have \vi\ colors of less than 0.5 and thus sit in the lower left corner of
Fig.\ 7b.  While three galaxies do not make a trend, it is not unexpected
that the faintest objects---and thus the most quiescent and metal-poor
ones---would be the bluest.  While our survey has not selected against
galaxies undergoing a recent starburst (e.g., blue compact dwarfs), the
lower right-hand corner of the plot is not filled.  This implies that
high-surface-brightness dwarfs are considerably rarer than LSB ones
(\cite{sch97}).  The general lack of a trend for the faintest galaxies to be
the reddest in \vi\ is an additional piece of evidence that these dwarfs are
not faded once-bright galaxies.  McGaugh \& Bothun (1994) reached the same
conclusions for the similar lack of correlation seen for larger LSB
galaxies.

\section{Summary and Conclusions}

	Of an optically-selected sample of 112 low-surface-brightness
galaxies with good optical ($V$ and $I$ band) and single-dish HI (Arecibo)
detections, we have chosen 101 to form the dwarf galaxy sample for this
work.  All 101 have heliocentric velocities greater than 500 km s$^{-1}$ and
absolute $V$ magnitudes fainter than $-19$.  The dwarfs are separated into
two types by their diameter at 25 $I$ mag arcsec$^{-2}$ (D$_{25}$): 55
``true dwarfs'' with diameters less than 7.5 kpc and 46 ``Magellanic
dwarfs'' with diameters greater than 7.5 kpc.  In general, this division
also separates galaxies with smaller scale lengths, lower optical
luminosities, and smaller neutral gas masses from their larger cousins.

	These low-surface-brightness dwarfs follow basically the same
trends seen in larger low-surface-brightness galaxies.  They are
quite blue in \vi, with an median value of 0.75.  The smaller ``true dwarfs''
have a median \vi\ color 0.1 magnitudes bluer than the Magellanic
dwarfs.  There are no correlations between the disk parameters $\alpha$
(scale length) and $\mu_0$ (central surface brightness), or between
these parameters and \vi, with the exception of the size/color trend
(since $\alpha$ is a size measurement) mentioned above.

	These dwarfs are also rich in neutral hydrogen, with a median HI
mass to $V$ luminosity ratio of 2 in solar units.  The median ratio
is somewhat higher for true dwarfs than for Magellanic dwarfs.  These high
ratios, along with the blue \vi\ colors and low surface brightnesses of the
dwarfs, indicates that the galaxies in our sample are likely to have very
low metallicities and little past star formation.  They are {\it not}
once-bright galaxies that have faded to obscurity, but rather quiescent
objects that have undergone little star formation in the age of the
universe.

	The current data set on these dwarf galaxies needs to be
supplemented.  The data presented in this paper implies that this sample
of dwarfs has undergone little past star formation, but $U$ and $B$ band
observations are required to truly understand the star formation history
of these objects.  Narrow-band emission-line imaging (e.g.,
H$\alpha$) would show where current star formation is occurring.
Interferometric HI observations would reveal not only the surface density of
neutral hydrogen in these galaxies, but could also be used to determine the
extent of the gas and the galaxies' dark matter content (via rotation
curves).  Small, faint field galaxies such as those found in our sample are
only beginning to be examined, and further study is likely to reveal much
about the evolution of this numerous class of objects.

\acknowledgements

We wish to thank the generous support of the Michigan-Dartmouth-MIT
Observatory for making this research possible.  We are also grateful
to the referee for suggestions that improved the presentation of this
paper.  R.A.P. was supported by a
National Science Foundation Graduate Fellowship and an understanding
graduate advisor during part of the research described in this paper.  This
research has made use of the NASA/IPAC Extragalactic Database (NED) which is
operated by the Jet Propulsion Laboratory, California Institute of
Technology, under contract with the National Aeronautics and Space
Administration.

\clearpage

\begin{deluxetable}{lccc}
\tablewidth{0pt}
\tablecaption{Observing runs}
\tablehead{\colhead{Dates} & \colhead{CCD} & \colhead{Binning} & 
\colhead{Resulting Pixel Size} }
\startdata
1992 Mar 20--26 & Thomson 400 $\times$ 576 & $1 \times 1$ & 0\farcs25
$\times$ 0\farcs25 \nl
1992 May 13--17 & Thomson 400 $\times$ 576 & $1 \times 1$ & 0\farcs25
$\times$ 0\farcs25 \nl
1992 Jun 17--22 & Thomson 400 $\times$ 576 & $2 \times 2$ & 0\farcs50
$\times$ 0\farcs50 \nl
1992 Nov 7--9 & Loral 2048 $\times$ 2048 & $3 \times 3$ & 0\farcs51
$\times$ 0\farcs51 \nl
1993 Mar 12--15 & Loral 2048 $\times$ 2048 & $3 \times 3$ & 0\farcs51
$\times$ 0\farcs51 \nl
1993 Apr 29--May 3 & Loral 2048 $\times$ 2048 & $3 \times 3$ & 0\farcs51
$\times$ 0\farcs51 \nl
1994 Feb 27--Mar 2 & Loral 2048 $\times$ 2048 & $3 \times 3$ & 0\farcs51
$\times$ 0\farcs51 \nl
\enddata
\end{deluxetable}

\clearpage

\begin{deluxetable}{lcccccccl}
\footnotesize
\tablenum{2}
\tablewidth{0pt}
\tablecaption{Basic properties of galaxies in initial sample}
\tablehead{\colhead{Name} & \colhead{$v_{hel,HI}$}&
\colhead{M$_V$} & \colhead{log(M$_{HI}$/M$_{\sun}$)} &
\colhead{R$_{25}$ (kpc)} & \colhead{$<\vi>$} & \colhead{$\alpha$ (kpc)} &
\colhead{$\mu_{0,I}$} & \colhead{Other Names} }
\tablecomments{The column $\mu_{0,I}$ is the central surface brightness of
the fitted exponential disk component in $I$ magnitudes arcsec$^{-2}$.
Quantities with subscript ``HI'' are calculated from data given in Eder et
al.~(1997) or from NED (NASA-IPAC Extragalactic Database).  All
distance-related values assume H$_0$=85 km s$^{-1}$ Mpc$^{-1}$,
$\Omega_0$=0.2, and a Virgocentric infall model with $v_{infall}$=300 km
s$^{-1}$ and D$_{Virgo}$=15 Mpc.  Dwarf spiral galaxies discussed in
Schombert et al.~(1995) are designated in the Other Names column by ``dS''.}
\startdata
D495-1  & 2309 & $-15.39$ & 7.95 & \phn2.3 & 0.75 & 0.7 & 21.5 & \nl
D495-2  & 2197 & $-15.48$ & 8.15 & \phn2.0 & 0.57 & 0.6 & 21.5 & \nl
D495-3  & 2267 & $-14.40$ & 8.09 & \phn3.2 & 0.96 & 0.8 & 22.5 & \nl
D500-2  & 1277 & $-16.43$ & 9.01 & \phn3.0 & 0.42 & 0.8 & 21.3 & UGC 5716 \nl
D500-3  & 1347 & $-15.77$ & 8.05 & \phn1.7 & 0.31 & 0.5 & 21.2 & \nl
D500-4  & 1576 & $-15.47$ & 7.74 & \phn1.6 & 0.70 & 0.4 & 20.3 & \nl
D500-5  & 4078 & $-18.23$ & 9.26 & \phn5.6 & 0.49 & 1.5 & 20.9 & \nl
D508-2  & 1924 & $-16.86$ & 8.82 & \phn3.6 & 0.31 & 2.1 & 23.0 & \nl
D508-5  & 4406 & $-17.05$ & 8.81 & \phn5.4 & 0.70 & 1.4 & 21.4 & \nl
D512-1  & 7199 & $-16.96$ & 9.33 & \phn3.8 & 0.65 & 1.4 & 21.9 & \nl
D512-2  & \phn831 & $-14.87$ & 8.02 & \phn1.6 & 0.80 & 0.5 & 21.1 & \nl
D512-3\tablenotemark{a}& 4201 & $-15.91$ & 8.73 & \phn3.1 & 0.89 & 1.2 & 21.9 & \nl
D512-4\tablenotemark{a}& 4805 & $-18.91$ & 9.37 & \phn7.1 & 0.74 & 1.7 & 20.1 & ZWG 163.075 \nl
D512-6  & 4269 & $-17.66$ & 9.43 & \phn6.5 & 0.91 & 2.0 & 21.4 & \nl
D512-7  & 4261 & $-16.90$ & 9.25 & \phn3.3 & 0.58 & 1.0 & 21.2 & \nl
D512-9  & 4499 & $-17.46$ & 9.17 & \phn4.8 & 0.83 & 1.6 & 21.2 & \nl
D512-10 & 4526 & $-16.33$ & 8.63 & \phn3.2 & 0.46 & 1.3 & 22.4 & \nl
D514-2  & 2023 & $-13.70$ & 8.54 & \phn1.2 & 0.23 & 0.7 & 23.2 & \nl
D514-5  & 6579 & $-16.45$ & 8.97 & \phn3.1 & 0.58 & 1.1 & 21.6 & \nl
D516-2  & 9767 & $-18.81$ & 9.33 & \phn8.7 & 0.72 & 2.3 & 20.9 & \nl
D516-3  & 9750 & $-16.68$ & 9.37 & \phn3.2 & 0.73 & 1.1 & 22.0 & \nl
D516-4  & 9322 & $-17.66$ & 9.07 & 10.0 & 1.33 & 3.3 & 21.7 & \nl
D561-2  & 4631 & $-16.49$ & 8.83 & \phn4.7 & 0.96 & 1.7 & 22.0 & \nl
D563-1  & 4161 & $-16.73$ & 8.54 & \phn3.6 & 0.44 & 1.5 & 22.2 & \nl
D563-2  & 4319 & $-17.81$ & 9.44 & \phn6.2 & 0.69 & 1.6 & 21.0 & F563-V2 \nl
D563-3  & 3935 & $-16.34$ & 8.74 & \phn4.1 & 0.83 & 1.7 & 22.3 & F563-V1 \nl
D563-4  & 3495 & $-17.08$ & 9.38 & \phn5.5 & 0.82 & 3.1 & 23.0 & F563-1; dS \nl
D563-5  & 4645 & $-17.36$ & 9.33 & \phn4.9 & 0.67 & 1.5 & 21.2 & \nl
D563-6  & 3567 & $-15.38$ & 8.19 & \phn2.1 & 0.88 & 0.9 & 22.0 & \nl
D564-2  & 8131 & $-16.88$ & 8.74 & \phn3.9 & 0.85 & 1.6 & 21.5 & \nl
D564-4  & 2703 & $-15.69$ & 8.23 & \phn2.4 & 0.87 & 0.8 & 21.4 & \nl
D564-8  & \phn488 & $-12.25$ & 7.21 & \phn0.4 & 0.93 & 0.4 & 22.8 & \nl
D564-9  & 3045 & $-18.00$ & 9.28 & \phn5.8 & 0.57 & 1.7 & 21.3 & UGC 4858 \nl
D564-11 & 4284 & $-14.52$ & 8.82 & \phn1.8 & 0.92 & 0.7 & 22.3 & \nl
D564-12 & 3840 & $-17.22$ & 9.10 & \phn3.8 & 0.70 & 1.4 & 21.3 & \nl
D564-13 & 4292 & $-16.64$ & 8.47 & \phn3.1 & 0.98 & 1.2 & 21.1 & \nl
D564-15 & 3386 & $-16.62$ & 8.70 & \phn3.7 & 0.78 & 1.1 & 21.2 & dS \nl
D565-1  & 6239 & $-16.94$ & 9.19 & \phn4.2 & 1.08 & 2.0 & 21.9 & \nl
D565-2  & 3830 & $-18.22$ & 9.56 & \phn6.5 & 0.56 & 2.2 & 21.7 & UGC 5005 \nl
D565-3  & 7623 & $-19.63$ & 9.33 & \phn6.1 & 0.76 & 1.6 & 21.2 & \nl
D565-5\tablenotemark{b}& \phn558 & $-13.39$ & 7.72 & \phn0.8 & 1.29 & 0.3 & 21.4 & \nl
D565-10 & \phn561 & $-12.82$ & 6.91 & \phn0.7 & 0.80 & 0.3 & 22.3 & KARA 68 056; F565-V4\nl
D568-1  & 1306 & $-18.95$ & 8.87 & \phn6.5 & 0.98 & 1.3 & 19.0 & UGC 5742; NGC 3287\nl
D568-2  & 1219 & $-14.24$ & 7.75 & \phn1.2 & 0.71 & 0.6 & 22.1 & \nl
D568-4  & 1557 & $-16.29$ & 8.25 & \phn2.7 & 0.58 & 0.7 & 20.7 & ZWG 094.005 \nl
D570-3  & 1383 & $-14.38$ & 7.77 & \phn1.5 & 0.42 & 0.6 & 22.2 & F570-V1 \nl
D570-4  & 1057 & $-14.52$ & 7.95 & \phn1.7 & 0.73 & 0.8 & 22.6 & F570-7 \nl
D570-6  & 1894 & $-14.86$ & 7.57 & \phn1.9 & 0.69 & 1.0 & 22.8 & \nl
D571-2  & 6796 & $-17.79$ & 9.19 & \phn6.1 & 0.82 & 1.9 & 21.3 & \nl
D571-5  & 6132 & $-16.59$ & 8.51 & \phn4.5 & 1.30 & 2.9 & 22.6 & \nl
D572-2  & 3746 & $-15.58$ & 8.61 & \phn2.3 & 0.34 & 0.7 & 21.8 & \nl
D572-4  & \phn793 & $-14.26$ & 7.75 & \phn1.0 & 0.53 & 0.4 & 21.6 & \nl
D572-5  & \phn976 & $-14.51$ & 8.13 & \phn1.3 & 0.52 & 0.4 & 21.2 & \nl
D575-1  & \phn580 & $-12.68$ & 7.24 & \phn0.7 & 0.70 & 0.2 & 21.8 & IC 3840; F575-2\nl
D575-2  & \phn764 & $-15.48$ & 8.71 & \phn3.0 & 0.87 & 1.2 & 22.3 & UGC 8011 \nl
D575-5  & \phn419 & $-11.35$ & 7.15 & \phn0.4 & 0.44 & 0.3 & 23.8 & KARA 68.215 \nl
D575-7  & 1013 & $-14.33$ & 8.16 & \phn1.3 & 0.65 & 0.4 & 21.5 & \nl
D576-3  & 6504 & $-18.54$ & 9.02 & 10.9 & 0.81 & 2.6 & 20.9 & \nl
D576-9  & 3831 & $-15.87$ & 8.87 & \phn3.1 & 0.53 & 1.2 & 22.3 & \nl
D577-2  & 6545 & $-16.82$ & 9.36 & \phn4.9 & 1.12 & 2.2 & 22.1 & \nl
D577-3  & 6936 & $-18.23$ & 9.06 & \phn6.3 & 0.84 & 1.8 & 20.9 & \nl
D577-5  & 4095 & $-16.82$ & 9.02 & \phn3.9 & 0.56 & 2.3 & 22.9 & dS \nl
D577-6  & 7572 & $-16.86$ & 9.11 & \phn4.9 & 0.89 & 1.9 & 22.3 & \nl
D582-2  & 4575 & $-15.12$ & 8.35 & \phn2.2 & 1.17 & 1.0 & 21.8 & \nl
D584-1  & 4669 & $-17.69$ & 9.11 & \phn5.2 & 1.02 & 0.9 & 19.7 & \nl
D584-2  & 2628 & $-16.12$ & 8.80 & \phn3.6 & 1.14 & 1.1 & 21.6 & \nl
D584-3  & 2576 & $-15.32$ & 8.71 & \phn2.9 & 1.08 & 0.8 & 21.3 & \nl
D584-4  & 2269 & $-16.35$ & 9.29 & \phn3.6 & 0.94 & 1.3 & 21.6 & \nl
D584-5  & 3104 & $-17.17$ & 9.51 & \phn4.4 & 0.76 & 1.4 & 21.2 & \nl
D584-6  & 4787 & $-17.01$ & 9.28 & \phn6.0 & 1.11 & 1.8 & 21.5 & \nl
D631-1  & 4698 & $-17.49$ & 9.11 & \phn5.0 & 0.67 & 1.4 & 21.1 & \nl
D631-7  & \phn343 & $-13.66$ & 7.81 & \phn0.8 & 0.55 & 0.3 & 21.3 & UGC 4115 \nl
D631-8  & 4471 & $-16.06$ & 8.83 & \phn3.0 & 0.73 & 1.0 & 21.7 & dS \nl
D634-3  & \phn318 & $ -9.76$ & 5.64 & \phn0.2 & 0.51 & 0.1 & 23.3 & \nl
D637-18 & 2740 & $-15.70$ & 8.67 & \phn3.6 & 0.94 & 1.1 & 21.6 & \nl
D637-20 & 6497 & $-17.70$ & 9.50 & \phn6.5 & 0.96 & 1.9 & 21.2 & \nl
D640-7\tablenotemark{c}& 1115 & $-14.37$ & 8.47 & \phn1.3 & 0.91 & 0.4 & 20.9 & UGC 5948 \nl
D640-11 & 6783 & $-18.18$ & 8.59 & \phn9.1 & 0.83 & 3.0 & 21.8 & \nl
D640-13 & \phn990 & $-13.70$ & 7.58 & \phn1.1 & 0.55 & 0.4 & 22.1 & F640-V1 \nl
D640-15 & 3244 & $-19.59$ & 9.29 & \phn8.7 & 0.93 & 1.6 & 19.0 & UGC 6211 \nl
D646-2  & 3580 & $-16.18$ & 9.24 & \phn3.0 & 1.07 & 1.1 & 21.2 & F646-3 \nl
D646-5  & 1044 & $-11.91$ & 8.17 & \phn0.8 & 1.60 & 0.3 & 21.6 & \nl
D646-7  & \phn213 & $-11.10$ & 6.40 & \phn0.4 & 0.84 & 0.1 & 21.6 & UGC 8091 \nl
D646-8  & 2101 & $-14.67$ & 7.43 & \phn1.4 & 0.11 & 1.6 & 24.1 & \nl
D646-11 & \phn565 & $-12.33$ & 7.05 & \phn0.6 & 0.48 & 0.2 & 22.3 & UGC 8061 \nl
D651-4  & 6299 & $-17.55$ & 9.24 & \phn6.2 & 0.77 & 2.2 & 21.7 & \nl
D656-1  & 4530 & $-17.68$ & 9.03 & \phn7.1 & 1.17 & 2.4 & 21.4 & UGC 10398 \nl
D656-2  & 1090 & $-15.08$ & 8.50 & \phn2.6 & 0.64 & 1.0 & 22.7 & UGC 10281 \nl
D656-5  & 8950 & $-16.44$ & 9.46 & \phn6.9 & 1.93 & 2.1 & 21.3 & \nl
D702-1  & 1879 & $-15.42$ & 8.60 & \phn2.1 & 0.67 & 0.6 & 21.3 & \nl
D704-2  & 4236 & $-17.31$ & 9.25 & \phn5.6 & 0.72 & 1.6 & 21.3 & \nl
D704-3  & 4174 & $-16.39$ & 8.87 & \phn3.7 & 0.63 & 1.6 & 22.1 & \nl
D709-5  & 9472 & $-17.61$ & 9.23 & \phn6.8 & 0.79 & 2.3 & 21.8 & KARA 68.060 \nl
D709-6  & 5228 & $-17.41$ & 8.48 & \phn6.0 & 1.03 & 1.6 & 20.8 & \nl
D709-10 & 2594 & $-14.73$ & 8.08 & \phn2.3 & 0.99 & 1.1 & 22.7 & \nl
D709-11 & 2993 & $-13.78$ & 7.90 & \phn1.7 & 1.23 & 0.9 & 22.6 & \nl
D721-5  & 5815 & $-16.95$ & 9.09 & \phn4.7 & 0.50 & 1.9 & 22.4 & F721-V4; dS \nl
D721-8  & 5433 & $-16.29$ & 9.06 & \phn5.4 & 1.61 & 1.7 & 21.5 & \nl
D721-10 & 1222 & $-14.19$ & 7.37 & \phn1.6 & 0.74 & 0.7 & 22.5 & F721-V2 \nl
D721-16 & 5260 & $-17.31$ & 9.03 & \phn4.8 & 0.62 & 1.4 & 21.2 & \nl
D723-3  & 7919 & $-16.25$ & 9.49 & \phn5.0 & 0.75 & 1.1 & 21.1 & \nl
D723-4  & 2166 & $-15.66$ & 8.92 & \phn2.6 & 0.61 & 0.9 & 22.0 & F723-V1 \nl
D723-5  & 1788 & $-16.44$ & 8.27 & \phn3.5 & 0.79 & 1.0 & 21.3 & F723-V2 \nl
D723-6  & 2092 & $-15.69$ & 8.74 & \phn2.5 & 0.37 & 0.8 & 21.7 & F723-1 \nl
D723-7  & 6922 & $-18.47$ & 9.67 & \phn7.4 & 0.58 & 2.4 & 21.4 & F723-2 \nl
D723-9  & 1690 & $-16.67$ & 8.83 & \phn4.4 & 0.71 & 1.4 & 21.6 & \nl
D774-1  & 4953 & $-19.07$ & 9.37 & \phn8.1 & 0.59 & 2.7 & 21.7 & dS \nl
D774-2  & 5555 & $-17.54$ & 9.28 & \phn4.9 & 0.50 & 1.6 & 21.3 & \nl
D774-3  & 7417 & $-17.33$ & 9.66 & \phn7.3 & 0.98 & 3.1 & 22.4 & \nl
\enddata
\tablenotetext{a}{D512-3 and D512-4 may be companions.}
\tablenotetext{b}{D565-5 is a rediscovery of UGC 5086, which was previously
lost.  It is a companion to NGC 2903.}
\tablenotetext{c}{D640-7 is a rediscovery of UGC 5948.}
\end{deluxetable}

\clearpage

\begin{figure}
\plotone{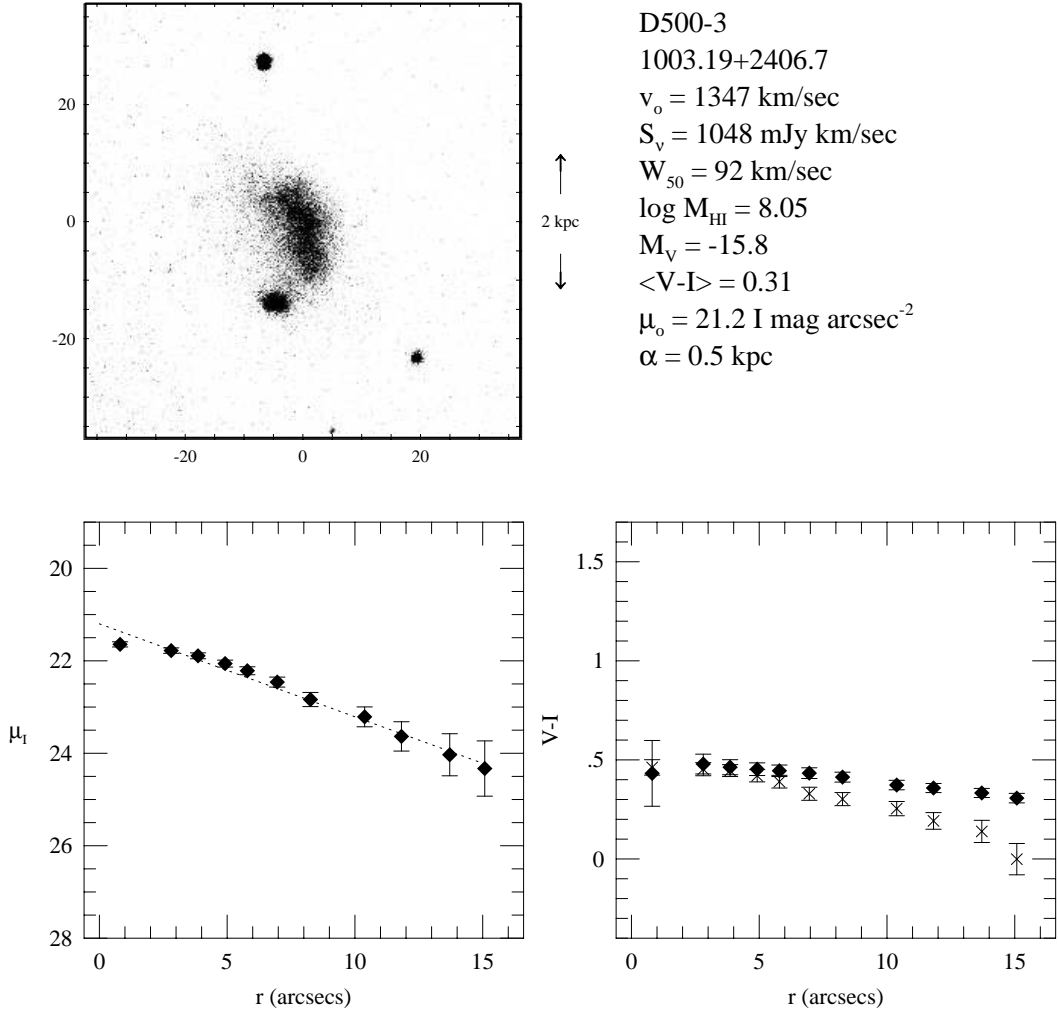}
\caption{Plots of color and surface brightness as a function of radius,
plus $I$ band grayscale pictures and basic facts for six representative
galaxies from our sample.  In the plots of \vi\ as a function of radius,
diamonds denote the aperture color within a particular radius and crosses
denote the differential color at that radius.  Similar plots for all the
galaxies in our sample will be available on the Dwarf Galaxy Survey homepage
(http://zebu.uoregon.edu/$\sim$js/dwarf.html).  (a) D500-3.}
\end{figure}

\clearpage

\begin{figure}
\figurenum{1b}
\plotone{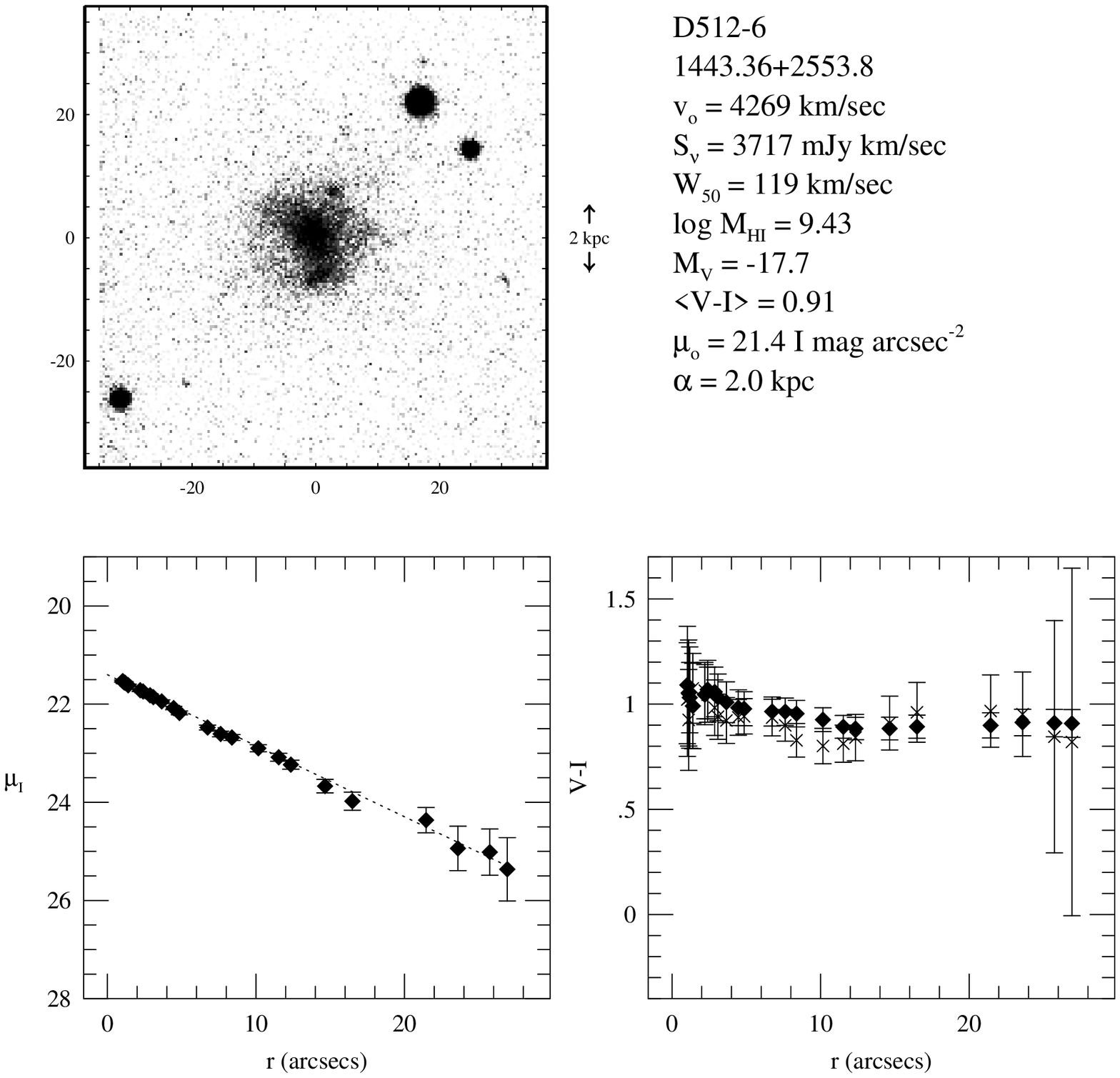}
\caption{D512-6.}
\end{figure}

\clearpage

\begin{figure}
\figurenum{1c}
\plotone{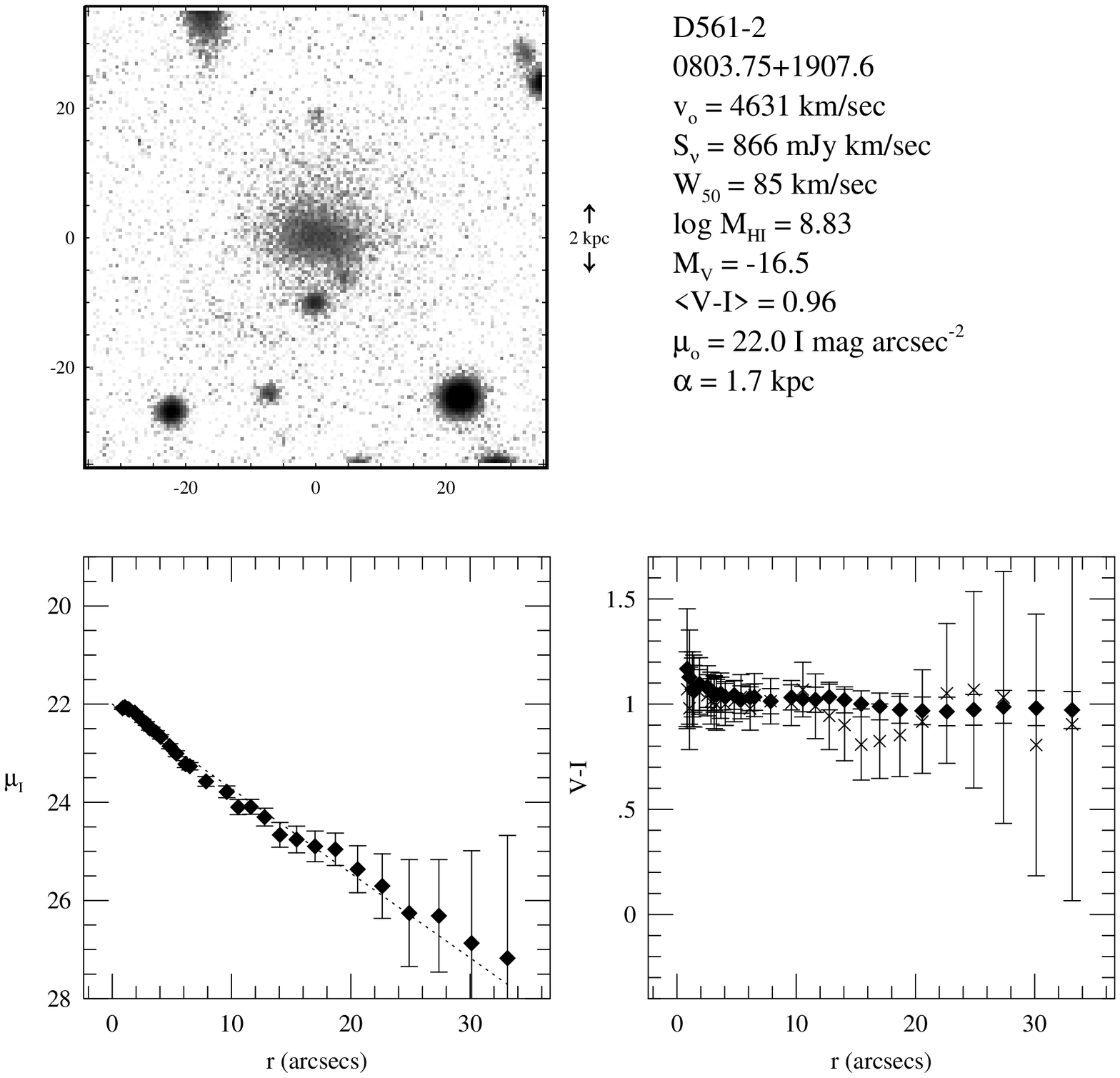}
\caption{D561-2.}
\end{figure}

\clearpage

\begin{figure}
\figurenum{1d}
\plotone{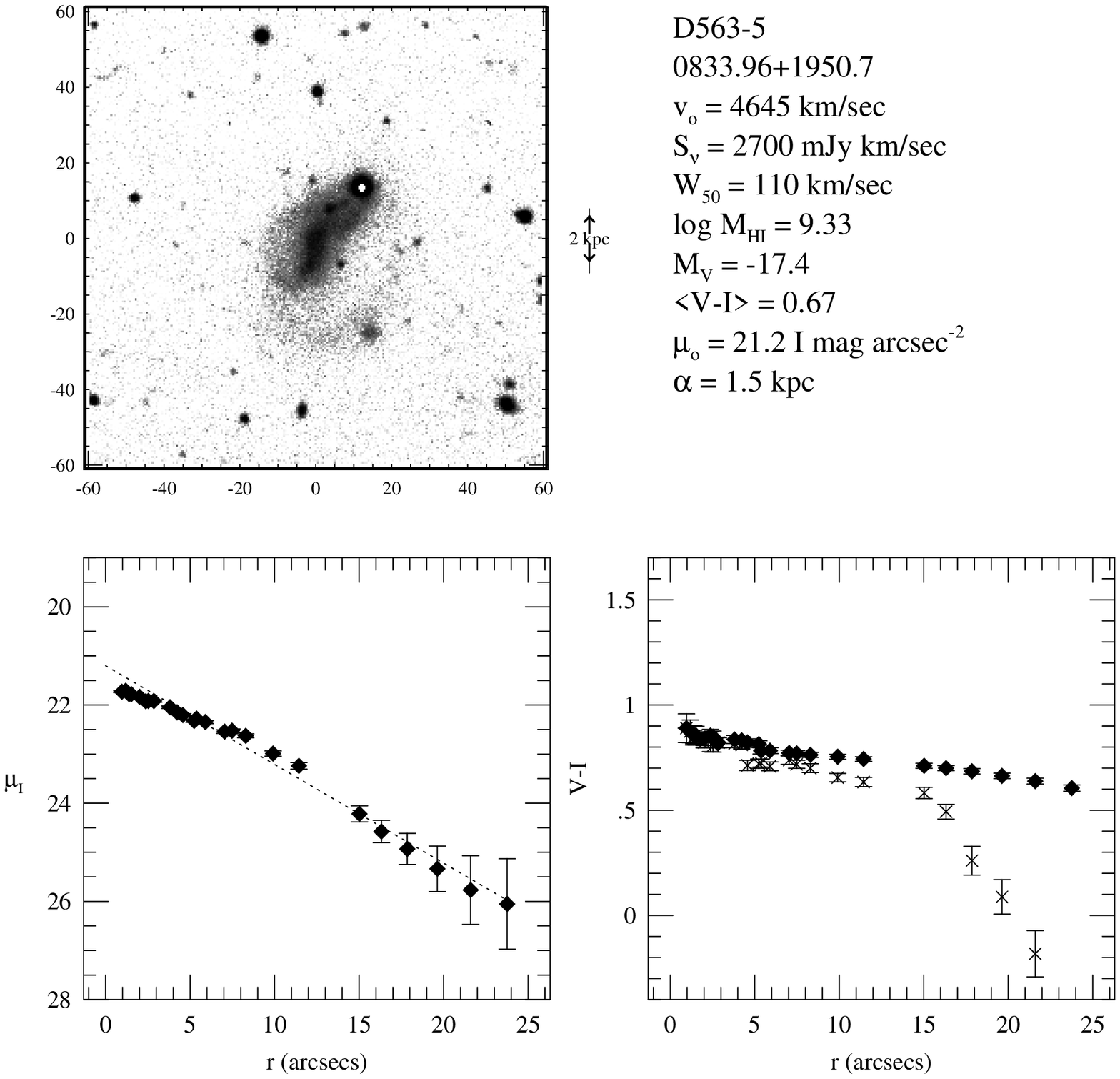}
\caption{D563-5.}
\end{figure}

\clearpage

\begin{figure}
\figurenum{1e}
\plotone{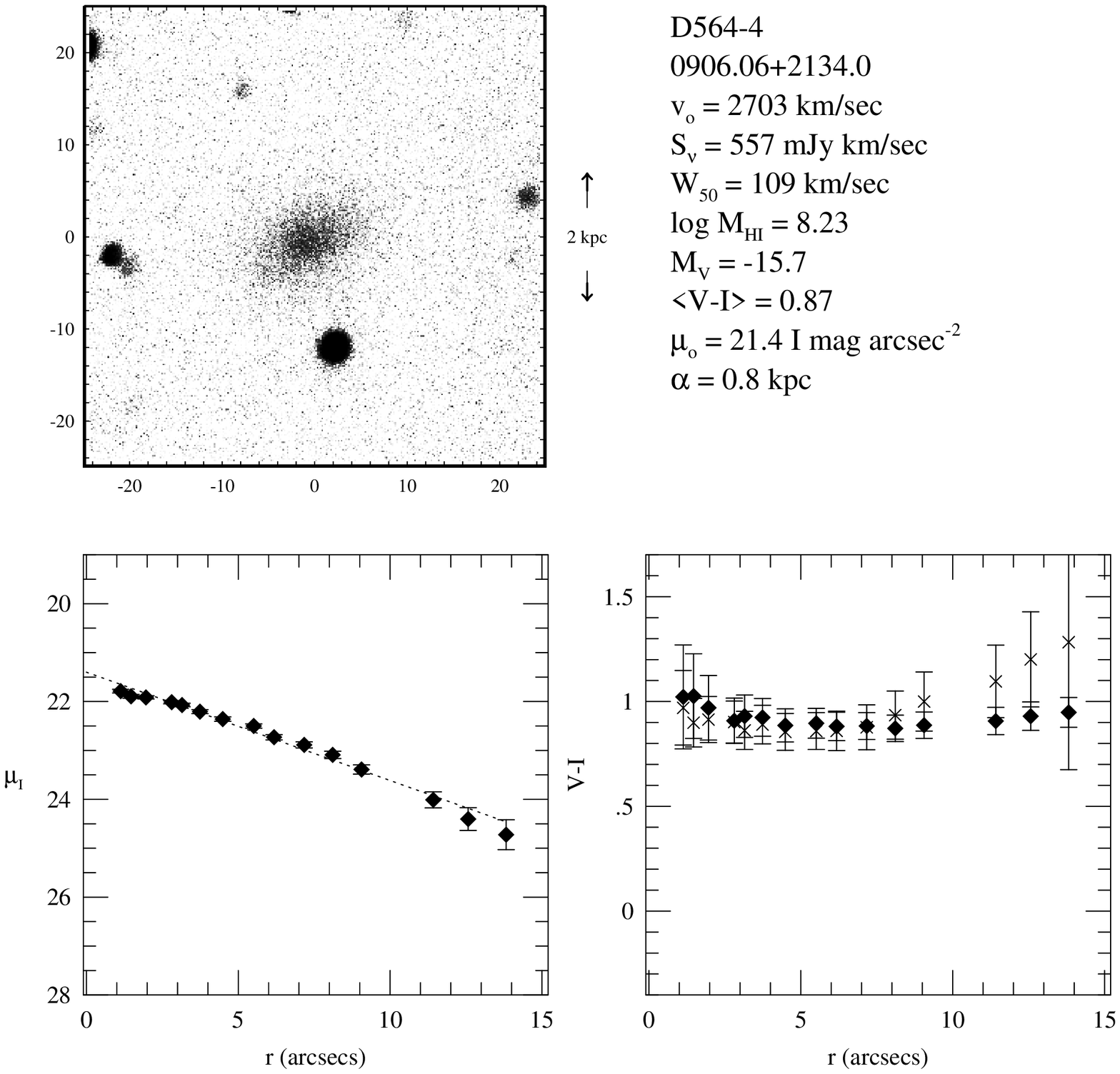}
\caption{D564-4.}
\end{figure}

\clearpage

\begin{figure}
\figurenum{1f}
\plotone{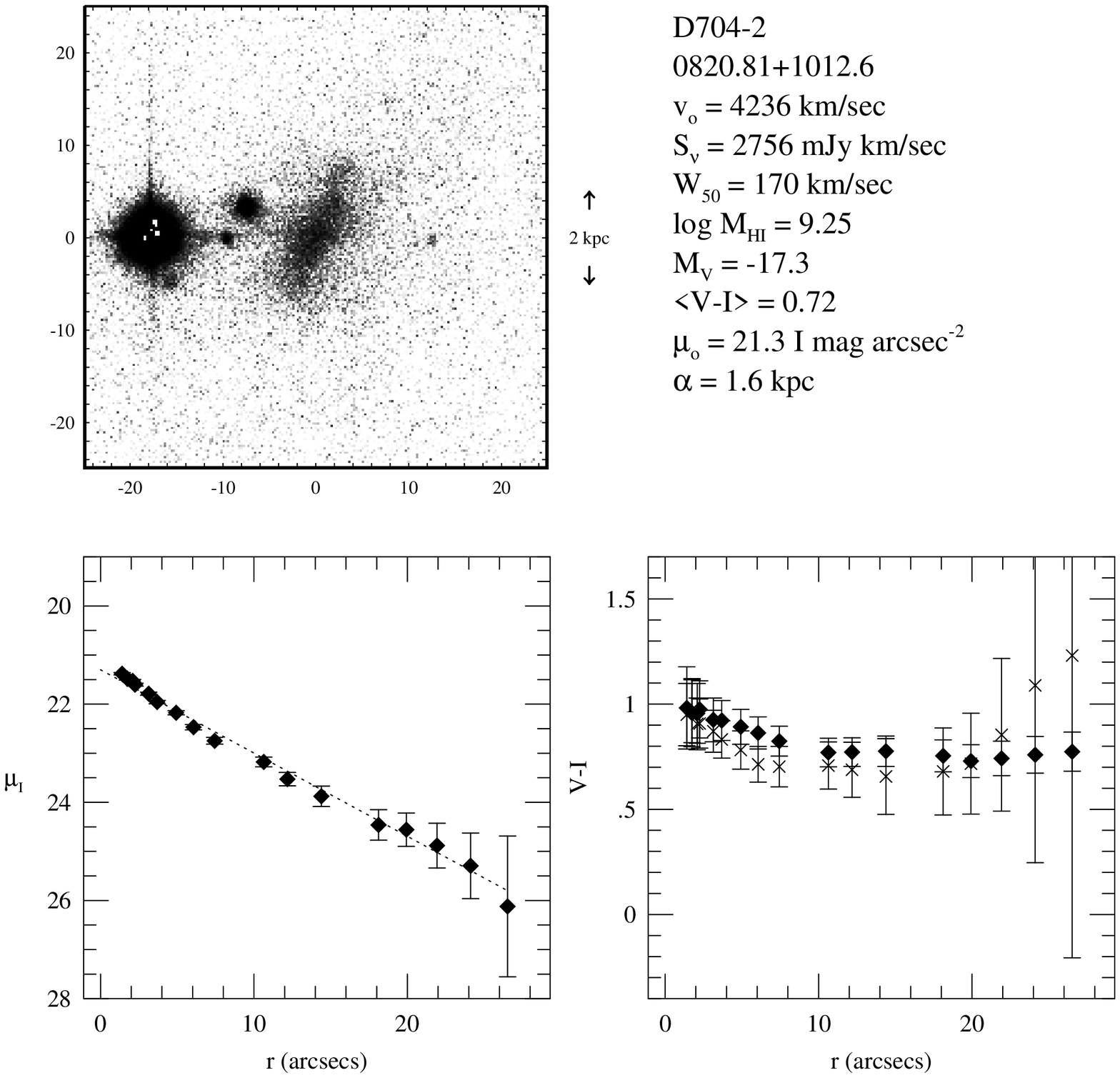}
\caption{D704-2.}
\end{figure}

\clearpage

\begin{figure}
\plotone{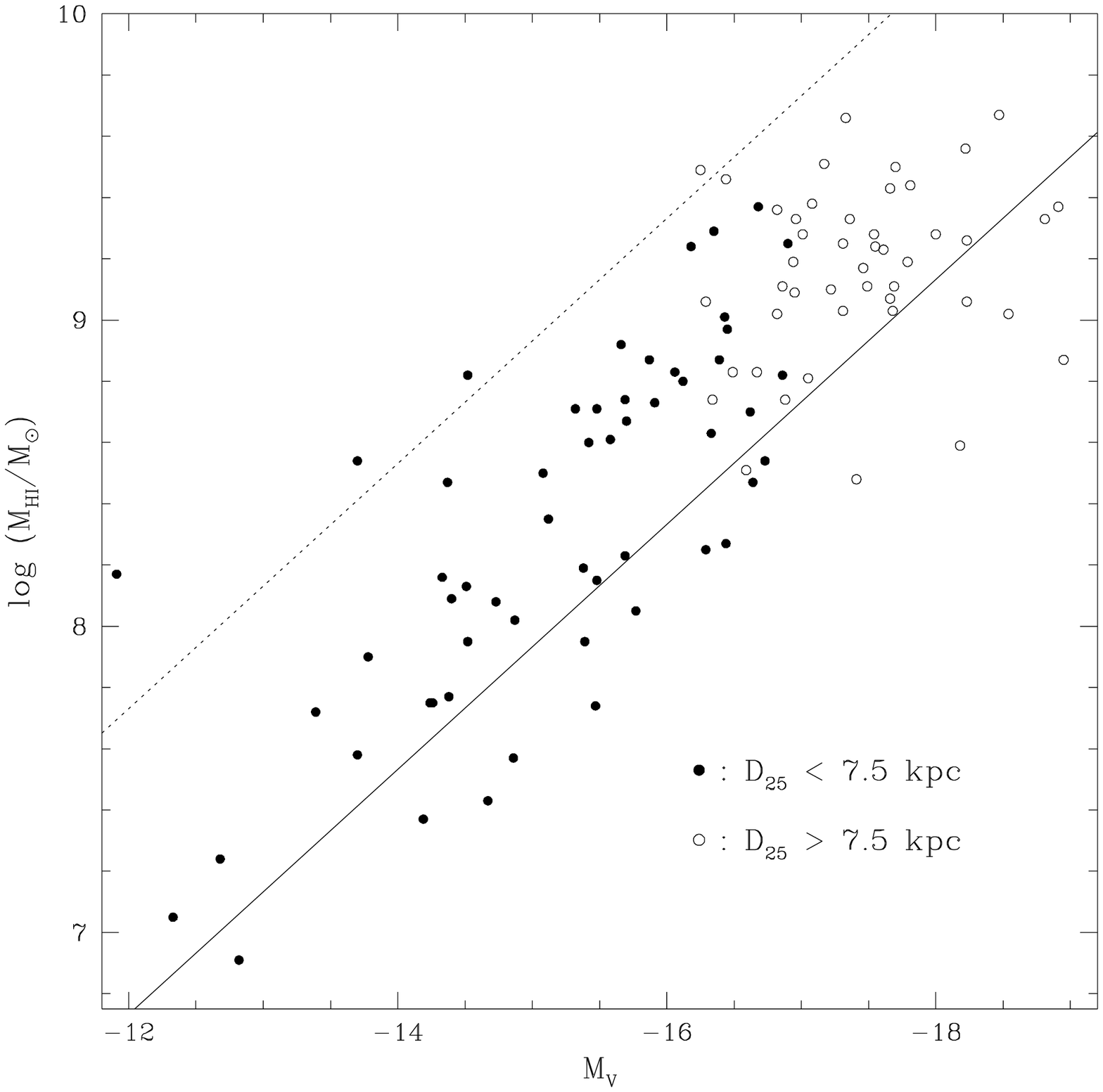}
\caption{The absolute $V$ magnitude of the dwarf galaxies in our sample
versus the log of their gas mass in solar masses.  The Magellanic dwarfs are
designated by open circles and the true dwarfs by filled circles.  The solid
line is the locus of points with M$_{HI}$/L$_V$=1 and the dotted line is
that for M$_{HI}$/L$_V$=10, where M$_{HI}$ and L$_V$ are in solar units.}
\end{figure}

\clearpage

\begin{figure}
\plotone{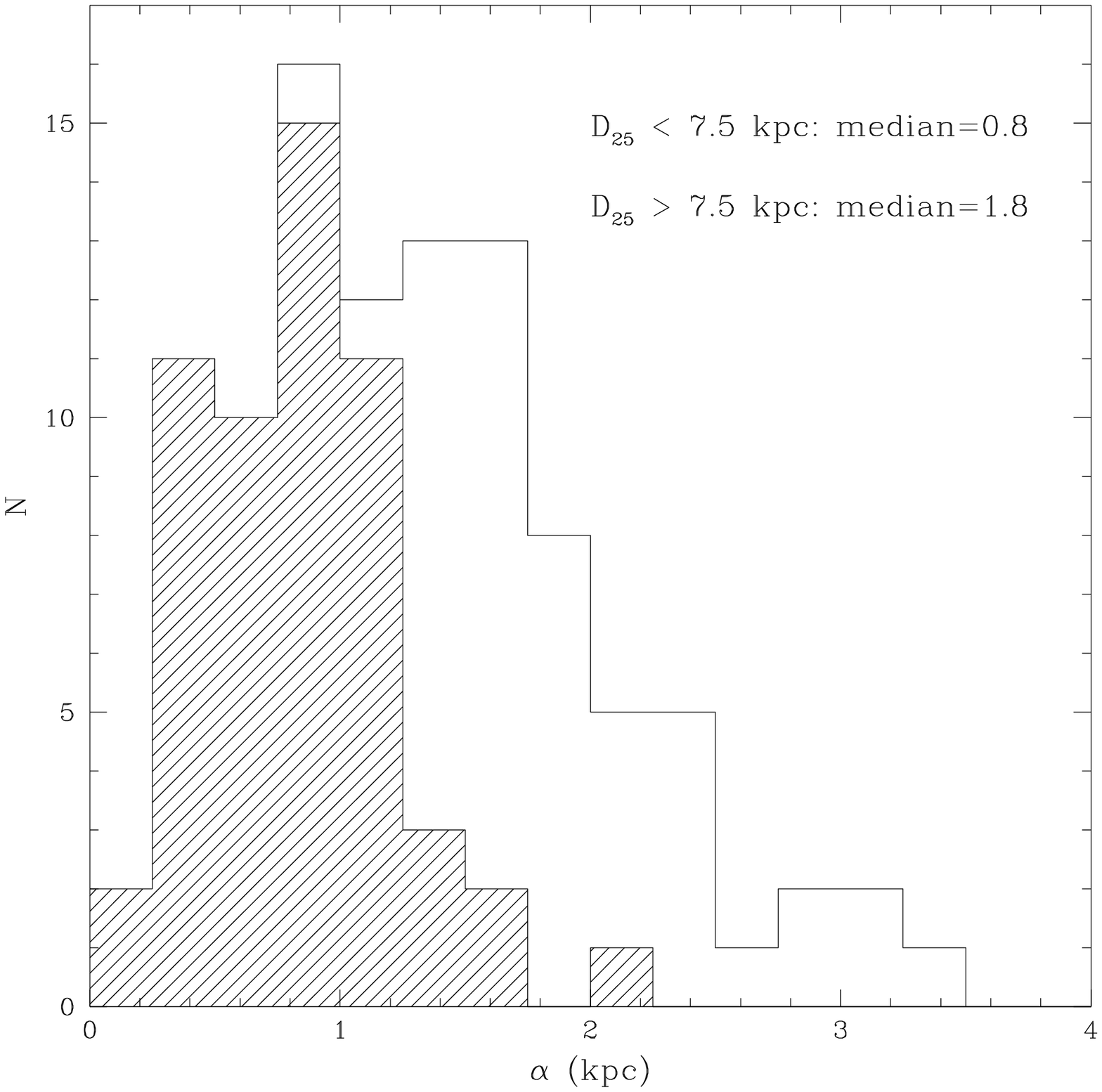}
\caption{The distribution of disk scale lengths of the dwarf galaxies in
our sample, with the true dwarf distribution shaded.  It is clear from this
histogram that the true dwarfs generally have smaller scale lengths than the
Magellanic dwarfs.}
\end{figure}

\clearpage

\begin{figure}
\plotone{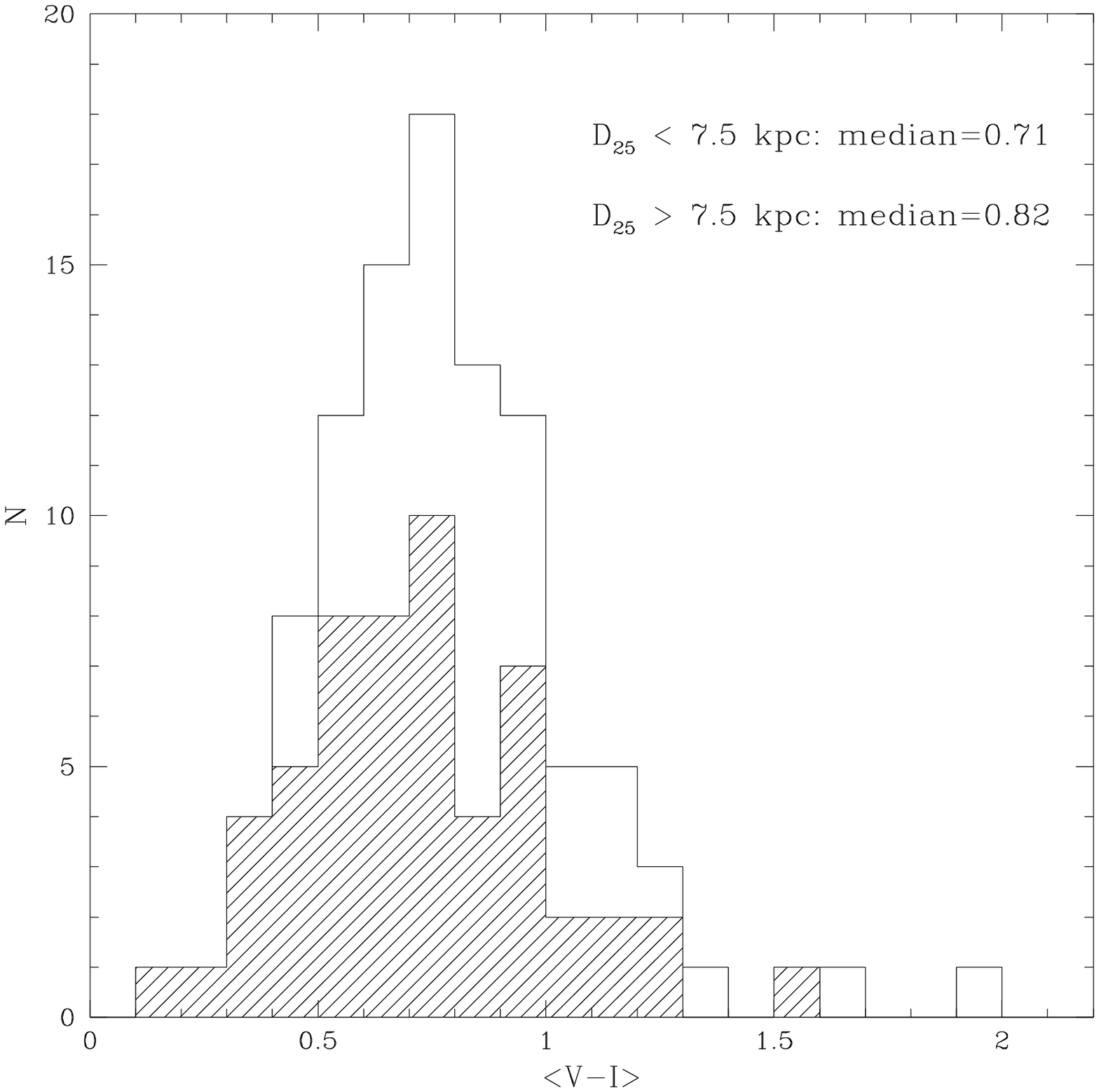}
\caption{The distribution of average \vi\ color in our sample of dwarfs,
with the true dwarf distribution shaded.  True dwarfs dominate the
blue tail, while Magellanic dwarfs dominate the red tail.}
\end{figure}

\clearpage

\begin{figure}
\plotone{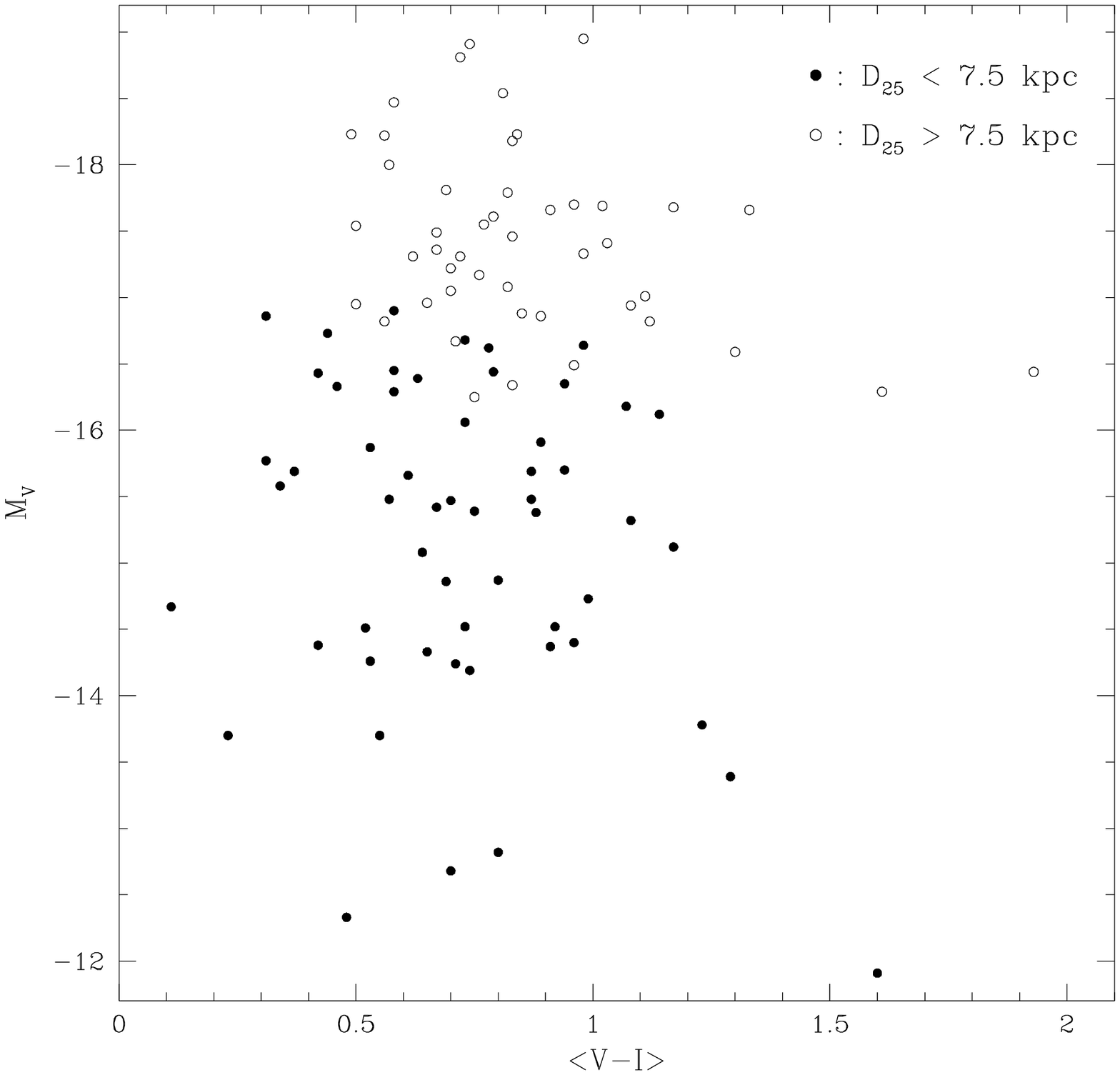}
\caption{Average \vi\ color of dwarf galaxies versus their absolute $V$
magnitude.  Magellanic dwarfs are marked with open circles and true
dwarfs with filled circles.  A slight trend for the less luminous galaxies
to be bluer than the more luminous ones is seen.} 
\end{figure}

\clearpage

\begin{figure}
\plotone{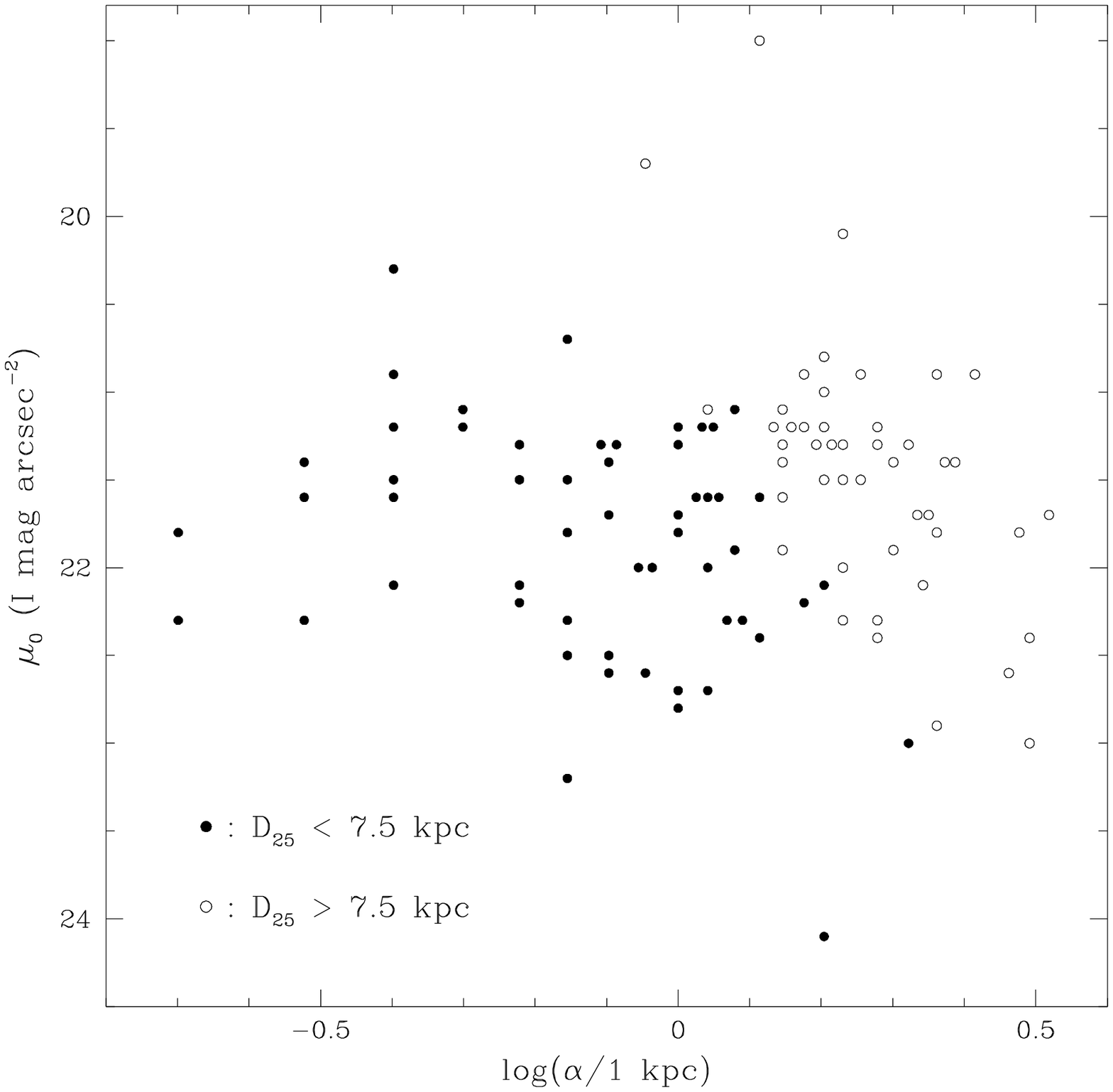}
\caption{The logarithm of the scale length of the disks of our dwarf
galaxies versus their disk central surface brightness.  The Magellanic dwarfs
are marked with open circles and the true dwarfs are marked with filled
circles.  No correlation is seen besides a selection effect against
small and very LSB objects.}
\end{figure}

\clearpage

\begin{figure}
\plotone{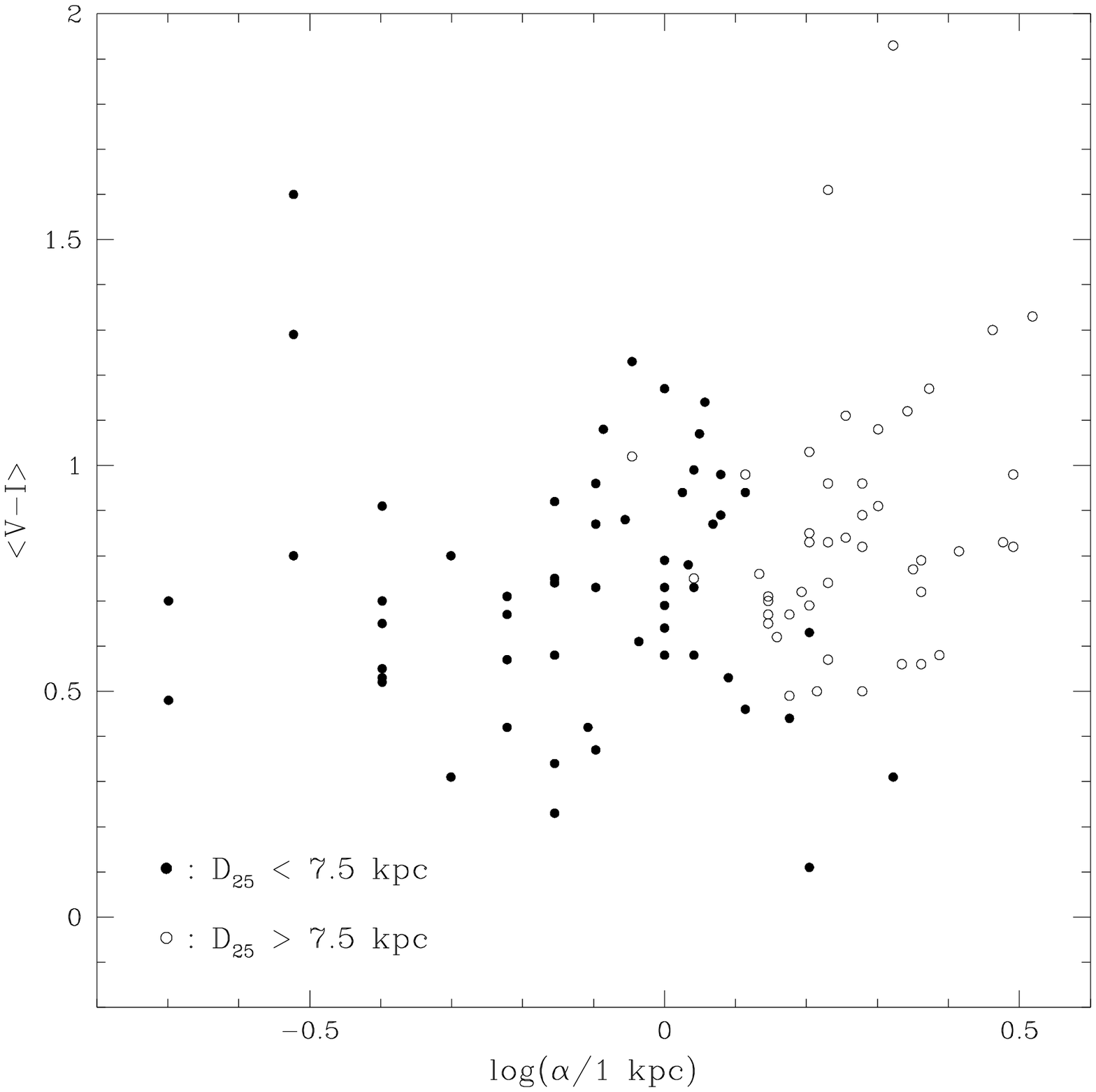}
\caption{Comparison of disk parameters with the overall \vi\ color
of the galaxies.  (a) The logarithm of the scale length $\alpha$ 
versus $<$\vi$>$.  True dwarfs are marked with filled circles; Magellanic
dwarfs with open circles.  On average, the smaller objects are bluer
than the larger ones.}
\end{figure}

\clearpage

\begin{figure}
\figurenum{7b}
\plotone{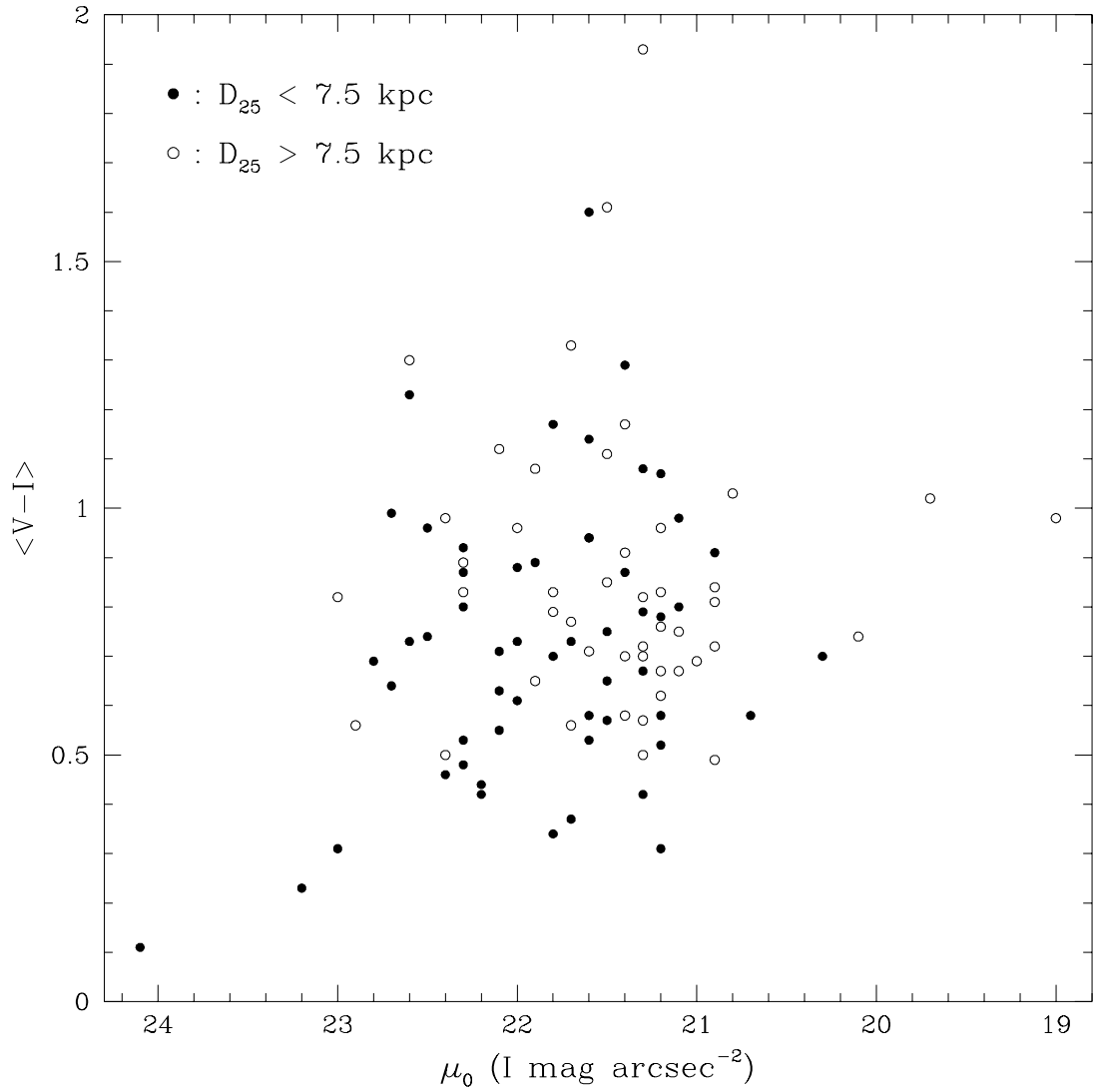}
\caption{The disk central surface brightness $\mu_0$
versus $<$\vi$>$.  Galaxies are marked using the same scheme as (a).
Little correlation is seen in this diagram.}
\end{figure}

\end{document}